# Navigating Chiral Spin Architectures in Non-Collinear Antiferromagnetic Thin Films

Freya Johnson[1,*], Jan Zemen[2], Geri Topore[3], Michele Shelly Conroy[3], Shanglong Ning[4], Jiahao Han[5,6], Shunsuke Fukami[5,6], Chiara Ciccarelli[1], Lesley F. Cohen[4]

[1] Cavendish Laboratory, University of Cambridge, Cambridge, CB3 0HE UK

[2] Faculty of Electrical Engineering, Czech Technical University in Prague, Technická 2, Prague, 160 00 Praha 6 Czech Republic

[3] Department of Materials and London Centre for Nanotechnology, Imperial College London, Exhibition Road, London SW7 2AZ, U.K.

[4] Department of Physics, Blackett Laboratory, Imperial College London, London SW72AZ, UK

[5] Laboratory for Nanoelectronics and Spintronics, Research Institute of Electrical Communication, Tohoku University, Sendai, Japan

[6] Center for Science and Innovation in Spintronics, Tohoku University, Sendai Japan

*Corresponding author, email: fj249@cam.ac.uk

## Abstract

Noncollinear antiferromagnets offer much promise for antiferromagnetic spintronics and neuromorphic applications with a plethora of functional properties surpassing many competing magnetic systems. Films grown on mismatched substrates may relieve strain by the creation of slip-plane defects that strongly manipulate global physical properties important for application. This work demonstrates that a post growth annealing strategy results in near-defect-free, structurally robust films that reveal the underlying thermal evolution of the magnetic order symmetry. Beyond a critical film thickness, the spin structure transitions between two right-handed irreducible chiral representations via left-handed chiral ordering producing a striking change in the sign and angular dependence of the Anomalous Hall coefficient. The previously established mechanism of spin rotations in the (111) plane cannot fully explain the transition in applied magnetic field, and using a macrospin model this work finds that rotations along the chirality-inverting [1-10] direction are energetically preferable under certain conditions. These observations suggest that both left-handed and right-handed chiral order can be accessed in a single system, opening new routes to engineer devices by controlled switching of magnetic chirality allowing selection of associated functional properties as governed by symmetry.

## Main

Non-collinear antiferromagnetic (nc-AFM) materials present a new and highly engaging frontier for spintronics, with proposed application in next-generation, energy-efficient “beyond von-Neumann” computing technologies [1-3]. Unlike the majority of collinear antiferromagnets, nc-AFM possess properties reminiscent of ferromagnetic materials, such as large anomalous Hall (AHE), anomalous Nernst (ANE) and magneto-optical Kerr effects (MOKE), despite having vanishing magnetization. Several collinear antiferromagnets have been identified as altermagnets that for different fundamental symmetry reasons may offer similar functional properties [4-6]. Critical for application in data storage is the development of ultra-thin, highly crystalline films and heterostructures, and the manipulation of their functional properties such as AHE and coercive field, $H_c$. It has previously been shown that the properties of nc-AFM thin films can be tuned by strain [7-9] and controlled by chemical doping [10], but engineering these may have undesirable impact on the crystallinity thereby suppressing intrinsic properties of interest [11].

As the magnetic order symmetry controls the boosted functional properties (related to the magnitude of the integrated Berry curvature), one topic of critical importance for nc-AFMs is the nature of the transformation of the magnetic order between the two possible irreducible representations (IR) and the two chiralities, represented schematically in Figure 1 [12]. Figure 1(a,c) represents the two IR with right-handed chirality – termed $\Gamma^{4g}$ and $\Gamma^{5g}$ respectively, while Figure 1(d,f) represents their counterparts with left-handed chirality, analogous to hexagonal type-B and type-A phases [13]. Both $\Gamma^{4g}$ and $\Gamma^{5g}$ have been observed experimentally in the cubic $Mn_3X$ materials ($X$ = Pt, Ir, Rh) and Mn antiperovskites $Mn_3AN$ ($A$ = Ni, Ga, Sn and many others), while type-A is the magnetic order of the relaxed hexagonal nc-AFMs $Mn_3Y$ ($Y$ = Sn, Ge, Ga) phase and type-B order has been reported to manifest in $Mn_3Sn$ thin films under tensile strain [14]. $\Gamma^{4g}$, $\Gamma^{5g}$ and Type-A support large longitudinal and transverse spin currents (in some cases without the requirement for spin-orbit coupling) [15]. These spin currents are shown to be essential to electrically manipulate magnetization in adjacent magnetic domains via a spin-transfer torque [15-19] or in the same domain via a self-induced torque [17,20]. Recent progress has shown that the same nc-AFMs host large tunnel magnetoresistance [19,21-24]. The specific components of the spin-conductivity tensors that are allowed depend on the magnetic order and the crystal directions considered - as an example, in a (001)-oriented cubic sample the longitudinal components $\sigma_{xx}^{x}$, $\sigma_{yy}^{y}$ and $\sigma_{zz}^{z}$ vanish for $\Gamma^{5g}$ but are allowed in $\Gamma^{4g}$ [25]. Importantly, $\Gamma^{4g}$ , type-A and type-B orders possess the symmetry-breaking allowing for AHE, ANE and MOKE while the $\Gamma^{5g}$ phase does not [26]. And finally, the magnetic order also impacts the small net moment which develops in these materials due to spin-canting because of either residual in-plane strain or the Dzyaloshinskii–Moriya interaction, facilitating domain manipulation [27-29]. For all these reasons, understanding how to transform spin order efficiently between different magnetic states will certainly yield better control over physical properties and unlock new functionalities.

The two IR $\Gamma^{4g}$ and $\Gamma^{5g}$ are related via a 90° simultaneous rotation of spins within the Kagome plane, allowing us to further define an intermediate state where the spins are rotated by an angle $\theta$, which is shown for the right-handed and left-handed cases in Figure 1(b) and (e), respectively. A full 180° rotation of spins in this plane does not change the IR or the chirality, but represents a switching of the antiferromagnetic order, and it is this switching that is most often considered for device application. Although the term “chirality” ordinarily implies two objects that are not related by a simple rotation, in this case transformation between left-handed and right-handed chirality is in fact possible, by rotations out of the Kagome plane [12]. But to our knowledge this has not been demonstrated experimentally. An ideal candidate to explore rotations between these topologically non-trivial magnetic orders is $Mn_3NiN$, which is structured with a triangular arrangement of Mn in the (111) plane. In the bulk, it enters the $\Gamma^{4g}$ phase from the paramagnetic state via a first order phase transition at 266 K and shows the $\Gamma^{5g}$ phase below T ~ 170 K [30-32]. Between these two temperatures, the material is believed to exist in the rotated phase (shown in Figure 1 (b)).

In this study we deposit high-quality $Mn_3NiN$ thin films of varying thickness on (001)-MgO substrate using pulsed laser deposition. X-ray diffraction (XRD), shown in Figure 1(g, h), reveals epitaxial growth with clear Laue oscillations, from which we extract the *c*-lattice parameter and the distance over which the X-rays are coherently scattered – the coherent thickness. Similar coherent growth has been observed using DC reactive magnetron sputtering for both $Mn_3GaN$ and $Mn_3NiN$ on LSAT [16,33].The *c*-lattice parameter is expanded for the 2 nm film, because this film is strained globally throughout its volume, but the c-lattice parameter quickly relaxes towards the bulk value for all other films (Figure 1 (i)). Figure 1 (j-m) shows representative high-angle annular dark-field scanning transmission electron microscopy (HAADF-STEM) for the thickest annealed sample, along with chemical analysis from energy dispersive X-ray spectrometry. The film grows homogenously, with high crystallinity.

As we previously established, the properties of nc-AFM are highly sensitive to local strain produced by defects [11], both HAADF-STEM and 4D-STEM strain mapping were employed to examine this aspect in further detail over a wider field-of-view (Figure 2). The overall misfit between MgO and $Mn_3NiN$ is 7.88%, and at the interface between the film and the MgO substrate, a regular array of edge dislocations is observed to accommodate this strain (Figure 2 (a)) via the domain matching epitaxy mechanism [34]. This is also evident in the 4D-STEM (Figure 2 (b-e)), where row-wise averaging of the strain in the $x$ and $y$ directions, $\varepsilon_{xx}$ and $\varepsilon_{yy}$, shows a peak directly at the interface, relaxing within the first 8 nm, and beyond that fluctuating between +/- 0.25 %, implying greatly reduced defect density. This interfacial region of high strain is considerably thinner than for unannealed $Mn_3NiN$ films grown on $SrTiO_3$ substrates [11].

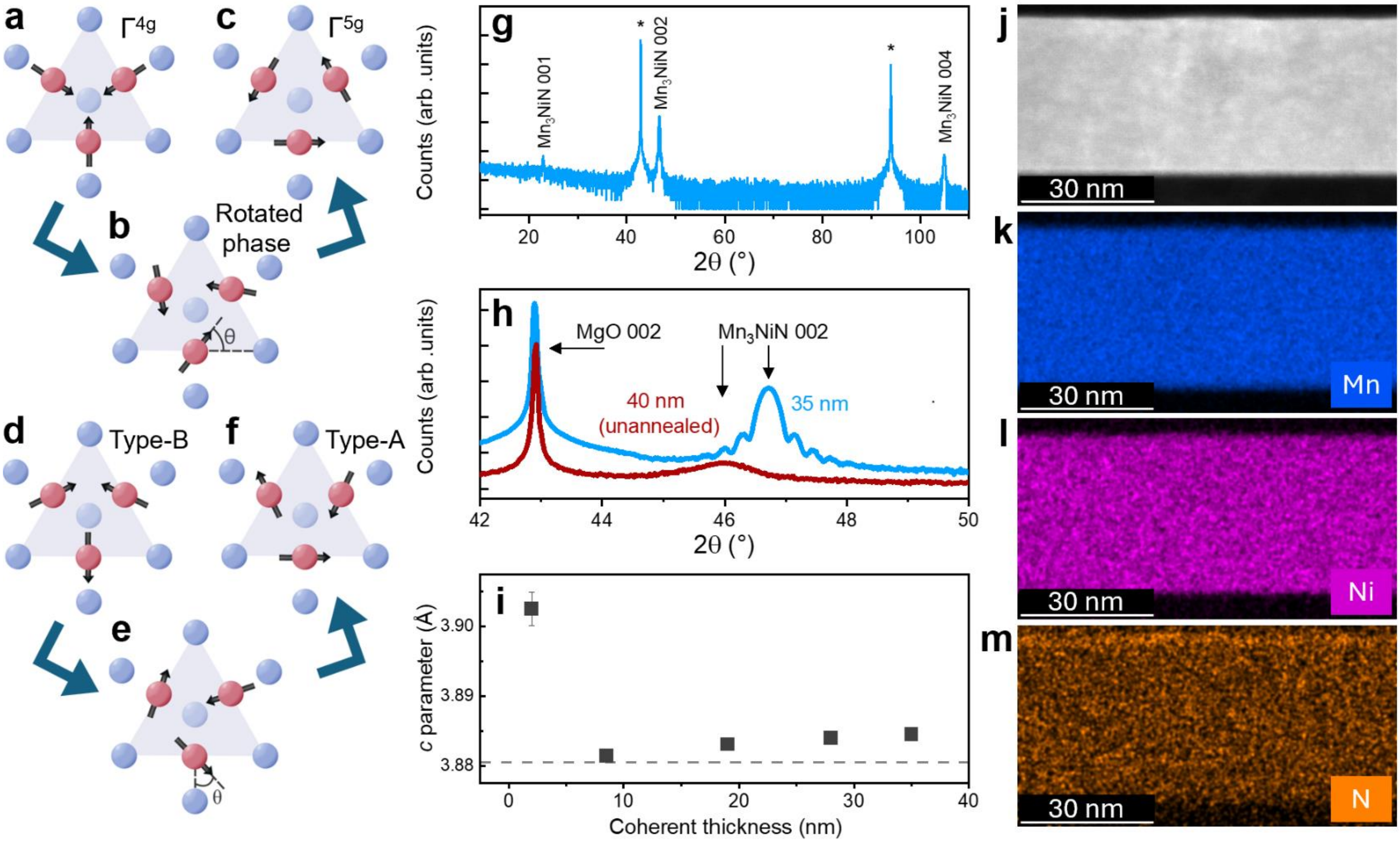

Fig 1. (a-e) Non-collinear antiferromagnetic structures. (a) The right-handed chiral irreducible representation $\Gamma^{4g}$ may transform to (c) the right-handed irreducible representation $\Gamma^{5g}$ via (b) the rotated phase. (d-f) The equivalent, but for left-handed chirality. (g-h) Typical X-ray diffraction from which we extract the out-of-plane lattice parameter and the coherent thickness. Asterisks indicate MgO substrate peaks. (h) Comparison between annealed and unannealed 002 film peaks. (i) *c* lattice parameter as a function of coherent thickness. The dashed line represents the lattice parameter of the $Mn_3NiN$ polycrystalline target. (j) High-angle annular dark-field image of thickest sample. (k-m) Energy dispersive X-ray spectroscopy showing compositional purity.

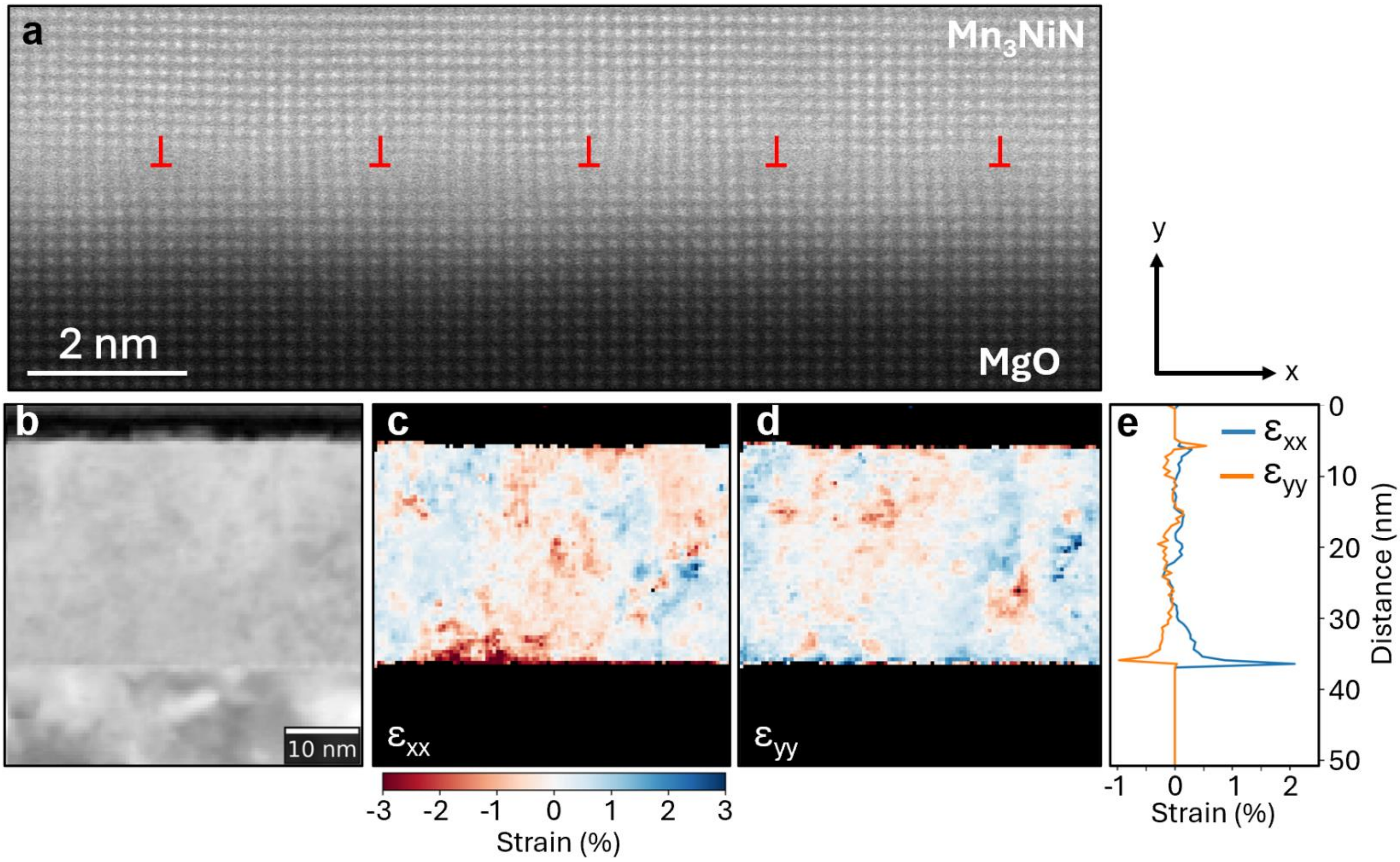

Fig 2. (a) Transmission electron microscopy image of the $MgO/Mn_3NiN$ interface revealing a regular array of dislocations. (b-d) 4D-STEM strain mapping of the thickest sample, in (c) x direction (d) y direction. Substrate and capping layer are masked in black. (e) Row-wise averaging of strain as a function of distance from the interface.

The electronic transport properties of these exceptional well-structured films are shown in Figure 3. Figure 3 (a) shows the longitudinal resistivity, $\rho_{xx}$, as a function of temperature for samples of different thickness. All samples exhibit a near-zero temperature coefficient of resistivity above 265 K. This effect, previously observed in this material family, has been attributed to strong magnetic scattering in the paramagnetic state above the Néel transition temperature, $T_N$ [33,35,36]. On cooling, the films exhibit metallic behavior, except for the 2 nm sample which shows a semiconductor-like temperature dependence, which we attribute to the higher strain in this sample and the influence of defects that we observe at the interface (see SI Figure 1). When increasing the coherent thickness, the metallicity is enhanced, before eventually saturating for the 28 and 35 nm samples. We achieve a residual resistance ratio RRR = $\rho_{xx}$ (280 K) / $\rho_{xx}$ (50 K) = 4.56 for the thickest sample, far higher than for unannealed films, [26] and also above high-quality films reported in the literature [16,37] consistent with the low defect density obtained from HAADF-STEM in our films.

Now we turn to the complex temperature and thickness dependence saturation value of anomalous Hall conductivity $\sigma_{xy,sat} = -\rho_{xy,sat}/\rho_{xx}^2$, where $\rho_{xy,sat}$ is taken from the high field 7 T data, as shown in Figure 3 (b). As mentioned previously, $\sigma_{xy}$ is forbidden by symmetry in both the strained and unstrained $\Gamma^{5g}$ magnetic state, but allowed in the $\Gamma^{4g}$ state, making transverse Hall measurements a useful tool to probe the spin order [26,38]. The films show no $\sigma_{xy}$ above ~265 K, corresponding to the paramagnetic phase seen in the temperature coefficient of resistivity. Comparing the 2 nm and 8.5 nm films, increasing the coherent thickness causes the peak value of $\sigma_{xy}$ to increase by a factor of ~20, which is mainly driven by the decrease in $\rho_{xx}$ by a factor of 3. This implies the defects at the interface observed in HAADF-STEM are a key source of electron scattering. In bulk samples, $\Gamma^{5g}$ forms below 170 K [30], but in the thinnest films the AHE survives far below this temperature. Further increasing the thickness to 19 nm causes a shift in properties. This film shows a sharp increase in $\sigma_{xy}$ below $T_N$, peaking at 250 K, before gradually decreasing and vanishing at 170 K. This behavior is consistent with measurements on bulk samples, indicating a second-order transition from $\Gamma^{4g}$ to $\Gamma^{5g}$ via the rotated phase. For yet thicker films an unexpected sign change is observed in $\sigma_{xy\text{-}sat}$ which cannot be explained by the current understanding of the transition. These films retain the key transition temperatures where the $\sigma_{xy}$ appears (at 260 K) and vanishes (at 170 K) consistent with the previous observations of $\Gamma^{4g}$ and $\Gamma^{5g}$ but in the interval between these temperatures the thicker films undergo an additional transformation not reported in inelastic neutron scattering experiments on bulk material samples where these symmetry transitions were originally set out in zero field [30-32,39]. Figure 3 (c-h) shows the full applied field (H) dependent transverse resistivity, $\rho_{xy}(H)$, for each film thickness at the same reduced temperature $T / T_N = 0.94$, where the sign change for the thickest films is evident.

The thickness dependence of the film properties and interest in the spin-lattice coupling in these materials stimulates further examination of the sign change region of the 35 nm coherent thickness film using temperature dependent XRD and magnetometry. Figure 3 (i,j) shows a clear correlation between the negative thermal expansion of the *c*-lattice parameter and the AHE in this region, confirming the relationship between spin order and lattice changes. In the bulk, the first order phase transition between the paramagnetic and non-collinear AFM phase is accompanied by a negative thermal expansion of the lattice [39-41] due to suppression of the local moment in the paramagnetic phase compared to the nc-AFM phase. We observe this behavior in both the 19 nm and 35 nm films, testifying to the single crystal-like quality of the samples (for comparative data on the 19 nm thick film please see SI figure 2). The magnitude of the lattice expansion appears similar in both films, but the 19 nm film has a considerably broadened transition. In zero-field cooled magnetometry, Figure 3 (k,l), we observe a magnetic transition at $T_N$ = 260 K in both in-plane and out-of-plane measurements.

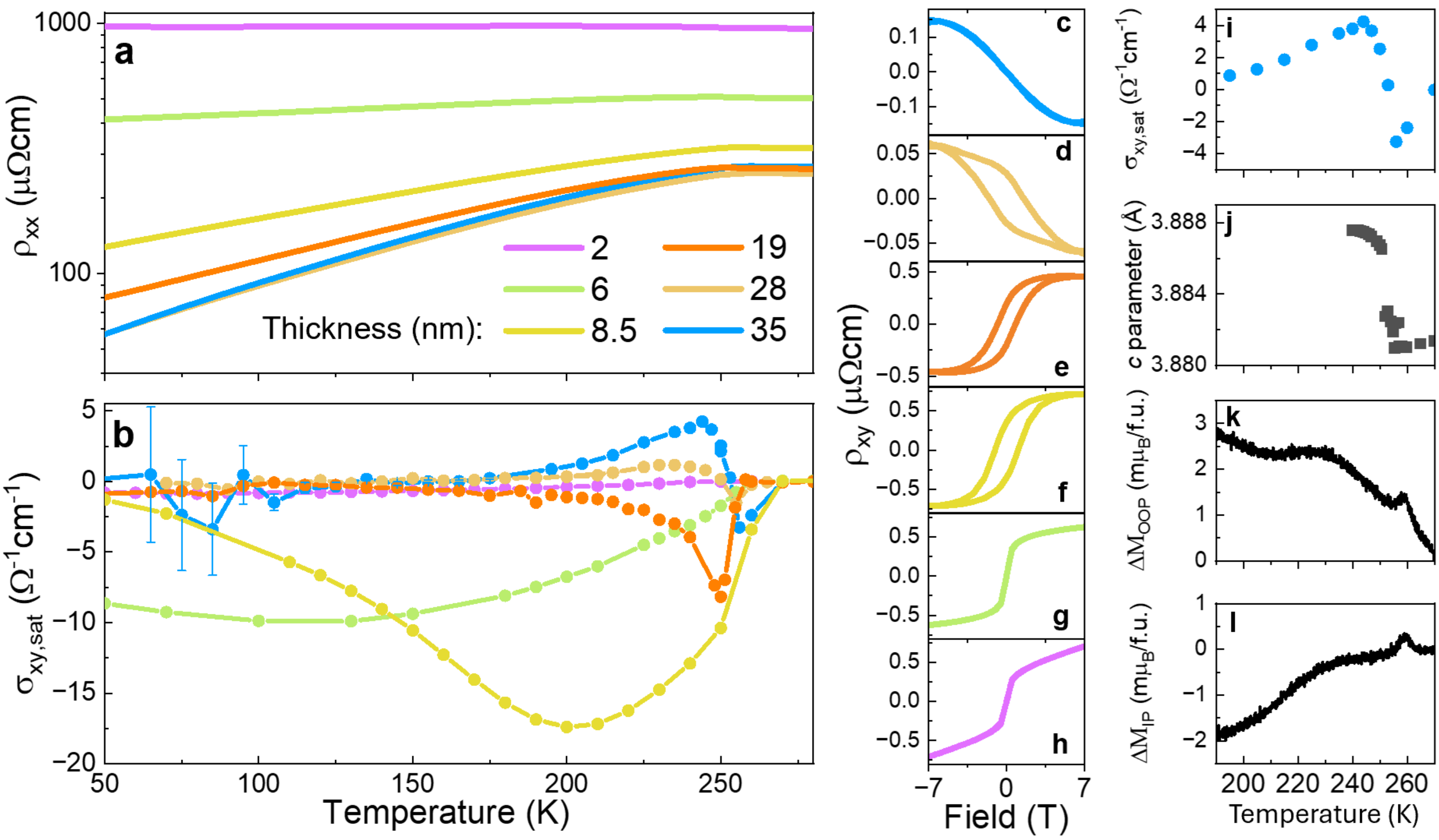

Fig 3. (a) Longitudinal resistivity as a function of temperature. (b) Saturated anomalous Hall conductivity $\sigma_{xy} = -\rho_{xy}/\rho_{xx}^2$, obtained by subtracting the linear ordinary Hall at high fields. (c-h) Anomalous Hall resistivity, $\rho_{xy}$, for all thicknesses at the same reduced temperature $T/T_N$ = 0.94. (i-l) Selected properties of the 35 nm sample. (i) Saturated anomalous Hall conductivity, (j) *c* lattice parameter extracted from X-ray diffraction for the 35 nm sample in the sign change temperature region. (k,l) Change of magnetization from paramagnetic phase in the same region in 50 mT applied field (k) out-of-plane (l) in-plane as measured using zero-field cooled magnetometry.

To probe the magnetic anisotropy of the sign change region for the thicker films, we patterned the 28 nm film into a 20-micron square Hall cross geometry and performed measurements in a rotation probe, varying the azimuthal angle $\varphi$. Figure 4a shows a schematic of the experimental configuration. We performed two sets of measurements – first, applying 7 T out-of-plane and sweeping $\varphi$ 360° (Figure 4 (b)), secondly by moving to specified angles and ramping field between +7 T and -7 T to measure a Hall loop in the usual way (Figure 4 (c-d)). We also measured the 19 nm and 8.5 nm samples in this way for comparison (Figure 4 (e-f)). We observe a markedly different angular dependence of AHE in the region before and after the sign change, confirming a striking change in the magnetic anisotropy and a transformation of the spin order in this region.

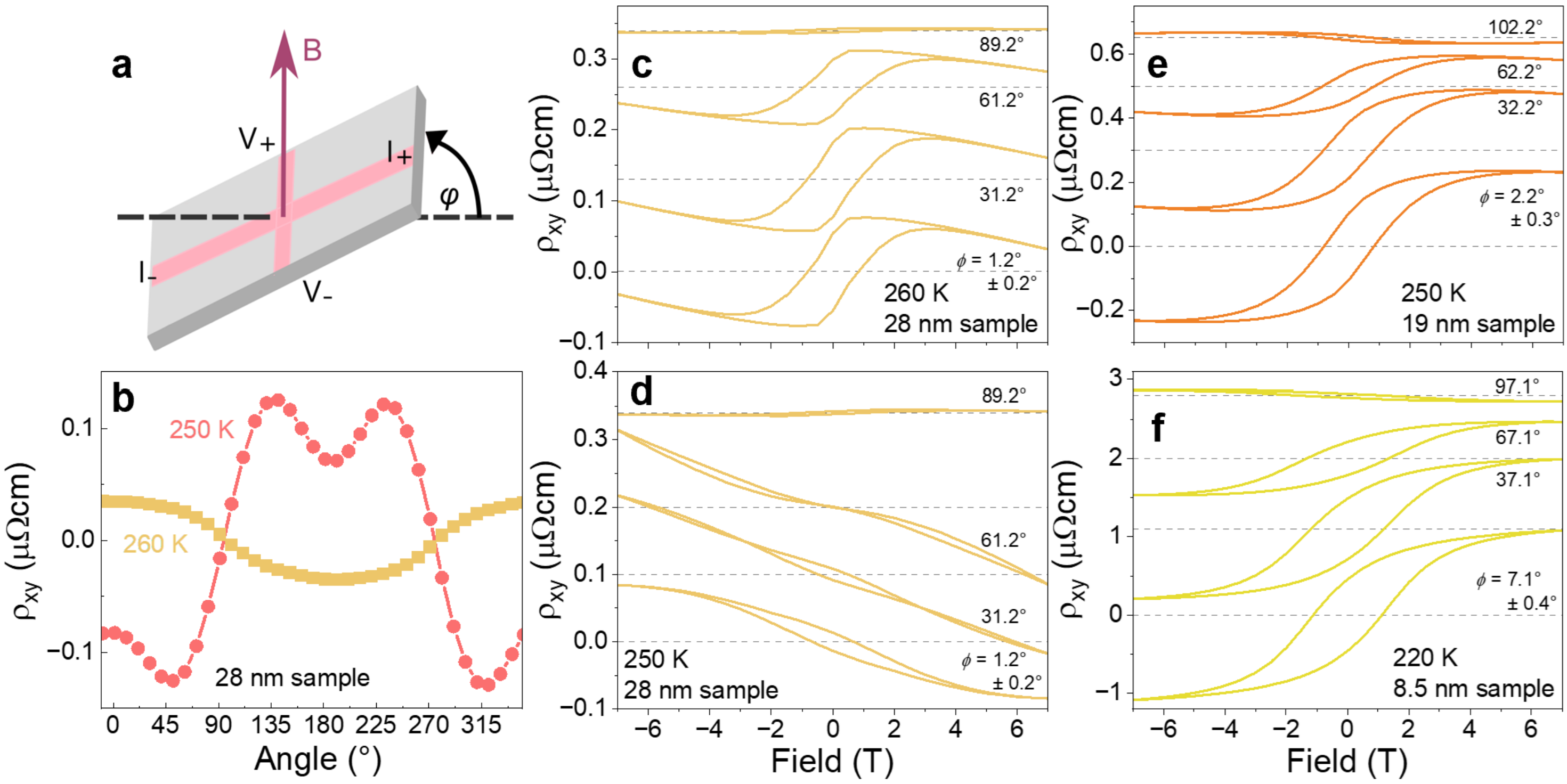

Fig 4. Measurements in a rotation probe. (a) Schematic of the measurement geometry. AHE in a Hall cross is measured as a function of azimuthal angle $\varphi$. (b) $\rho_{xy}$ measured in 7T while sweeping $\varphi$ through 360° for temperatures above and below the sign change. (c-f) AHE loops measured at select $\varphi$ and ramping magnetic field. Vertical offsets have been added for clarity. (c-d) AHE Loops for the 28 nm sample above and below the sign change temperature. (e-f) AHE Loops for the 19 nm and 8.5 nm samples.

## Discussion

In the following we set out various scenarios that might explain the sign change and angular dependence of the AHC for the thicker films, ruling out those that clearly are unfeasible.

**Change of sign of the net magnetization**: Observation of AHE in $Mn_3NiN$ and related nc-AFM in applied field is only possible due to the piezomagnetic effect [27,28,42] where strain can induce spin canting and hence produce a small net magnetization coupled to the antiferromagnetic order. This net moment is important in allowing the domain state to be manipulated, which would not be possible in a fully compensated state. As there are eight {111} planes and the $\Gamma^{4g}$ structure can be ordered in any of them [43], there are eight possible $\Gamma^{4g}$ domains, each with an associated net moment in one of the <112> directions. Applying a magnetic field in the positive z-direction preferentially couples to four of these eight domains with net moments oriented with the field, and these four domains all share the same sign of AHE [43]. It is known that changes in the electronic band structure can induce opposite direction of canting and opposite net moment while leaving the sign of AHE unaffected [29]. Therefore, at first glance, it is not implausible that the sign change for the thicker films is related to a small alteration of the canting in $\Gamma^{4g}$, which would lead to a reversal of the net moment. Such an alteration of the canting could be caused by a vertical strain distribution within the film, but our detailed 4D-STEM strain maps do not show evidence for this.

If the shape of the transverse voltage loop is dominated by domain re-orientation, then maximal signal should be obtained for the field oriented out-of-plane, when it efficiently orders all four phases with the same sign of AHE. This is indeed what is observed for the data at 260 K, Figure 4 (c), as well as for the 19 nm and 8.5 nm samples Figure 4 (e-f). But at 250 K maximal signal is obtained at $\varphi$ = 52° and 142°, which cannot be explained by a small change in the spin canting causing a reversal of the net moment in $\Gamma^{4g}$. For this reason, we can exclude this explanation.

**Domination of the Extrinsic AHE component**: A competition between intrinsic effects and extrinsic scattering/disorder contributions could also lead to a sign change, but extrinsic contributions are proportional to the longitudinal conductivity and therefore would increase in magnitude on cooling regardless of the intrinsic nc-AFM order. However, in Figure 3 (b) we see that $\sigma_{xy}$ in fact vanishes at lower temperatures, hence this scenario can also be ruled out.

**Changes in the Carriers / Band structure:** In semiconductors a change of carrier type from electron-like to hole-like or vice-versa is associated with a non-linear Hall effect which can resemble the anomalous Hall effect, but this effect is non-hysteretic unlike our observations. Changes to the carrier type and density as a function of temperature has been observed previously in both the hexagonal nc-AFMs and Mn antiperovskites. In $Mn_3GaN$, which adopts the $\Gamma^{5g}$ structure, a sign change in the ordinary Hall effect was observed at 200 K without any accompanying non-linear Hall signal [16]. Similar sign changes have been seen in Fe-doped $Mn_3Ge$ [44] without any associated change in the anomalous Hall. In $Mn_3Ge$, a large shift in carrier sign and concentration was observed over the 100 K – 5 K temperature range due to a Lifshitz transition, causing a change of the appearance of Hall resistivity $\rho_{xy}$ [45]. In our case, the effect in AHC we observe occurs over a narrow 10 K range at 250 K, and has a dependence on field angle, which prompts us consider alternative origins, although we cannot entirely rule out the possibility that some abrupt change in the electronic structure has occurred.

**Rotation of the non-collinear antiferromagnetic structure from $\Gamma^{4g}$ to $\Gamma^{5g}$**: The transition between $\Gamma^{4g}$ and $\Gamma^{5g}$ in bulk $Mn_3NiN$ has been understood phenomenologically in terms of a local uniaxial magnetic anisotropy ($K_u$) fixed in size and sign at each Mn site, but differing in direction, giving a sublattice-dependent axis where a negative sign favors $\Gamma^{4g}$ while a positive one favors $\Gamma^{5g}$ [46-48]. The change of sign of $K_u$ as a function of temperature necessarily implies that there is a temperature regime where the anisotropy becomes vanishingly small and at some temperature zero, and this is an interesting regime to examine in more detail. Calculations of the intrinsic AHC for rotations in the (111) plane have revealed a sinusoidal-like dependence on the rotation angle $\theta$, with a maximum at $\theta$ = 90° ($\Gamma^{4g}$) and a zero at $\theta$ = 0, 180° ($\Gamma^{5g}$) [49]. Rotating between $\Gamma^{4g}$ and $\Gamma^{5g}$ therefore cannot lead to a sign change of the AHC. Rotating continuously from $\Gamma^{4g}$ to $\Gamma^{5g}$ and then back to $\Gamma^{4g}$ can lead to a sign change due to a 180° rotation of the local moments, and was used to explain a sign change of $\sigma_{xy}$ in thin films of $Mn_3SnN$ [50]. However, this rotation would not change the strain-induced spin-canting and therefore a sign change of the net moment would be expected as explained in 1 above. Applying a field would then undo this change and not lead to a sign change in $\sigma_{xy}$.

**Alternative rotations for $\Gamma^{4g}$ transitions to $\Gamma^{5g}$**: For the above reasons, we conclude that the previously understood mechanism by which $\Gamma^{4g}$ transitions to $\Gamma^{5g}$ cannot apply here and instead explore alternative rotations other than in the (111) plane.

In the tight binding calculations by Zhou et al. [49] it was found that in $Mn_3NiN$ the left-handed spin chiralities can have favorable total energy to that of the right-handed versions, while the AHE is comparable in magnitude, and can have opposite sign. To explore further we build a macrospin model using parameters fitted to density functional theory (see methods). This model includes the uniaxial anisotropy term, $K_u$, and recreates the $\Gamma^{4g}$ ($\Gamma^{5g}$) energy minimum for negative (positive) anisotropy [46-48]. However, this model is insufficient to explain a smooth rotation of spins from $\Gamma^{4g}$ to $\Gamma^{5g}$, as it predicts an abrupt transition between the two states when the anisotropy changes from negative to positive. We modify this theory by introducing a small fixed tetragonal anisotropy term, $K_{tx} = K_{ty} \neq K_{tz}$ which resolves this insufficiency, and importantly we also include the effect of an external field along [001]. Figure 5 (a) shows this modified energy landscape.

As the sample crosses from $\Gamma^{4g}$ to $\Gamma^{5g}$, $K_u$ must vanish and therefore the small tetragonal anisotropy plays a role. In this case, we find rotations of the plane containing all three moments (the exchange plane) around the [1-10] axis does not cost significant energy, as all moments remain coplanar with this rotation and hence no exchange energy is involved. This rotation of the exchange plane results in a rotation of the vector chirality introduced by Szunyogh et al. in $Mn_3Ir$ [51], and can invert the handedness of the scalar spin chirality. This rotation is shown schematically in Figure 5 (b). When the applied field of 7 T is also considered, we find the left-handed ordering, analogous to Type-B, is energetically favorable over the right-handed $\Gamma^{4g}$ in this anisotropy region. We illustrate the evolution of the entire energy landscape on cooling using four cases: firstly, the situation just below $T_N$, where the uniaxial anisotropies are large and negative and right-handed $\Gamma^{4g}$ well known in bulk samples is preferred (Figure 5 (c)). Secondly, at the sign change temperature, with vanishing anisotropy and equal minima for right-handed $\Gamma^{4g}$ and left-handed $\Gamma^{4g}$ (Figure 5 (d)). Thirdly, just below the sign change temperature, with small positive anisotropies favoring left-handed $\Gamma^{4g}$ (Figure 5 (e)). Finally, far below, where large positive anisotropies lead to the rotation of spins in both the $\theta$ and $\varphi$ directions from left-handed $\Gamma^{4g}$ to the right-handed ordering $\Gamma^{5g}$, also well known in bulk samples at temperature below 170 K (Figure 5 (f)).

By mapping the energy landscape in this way, the technological implications of these previously overlooked in- field rotations become apparent. For example, a magnetic tunnel junction memory device that operates using two right-handed $\Gamma^{4g}$ states, rotated by 180° with respect to each other in the Kagome plane, cannot be switched using a spin torque that induces a rotation in the (111) plane – as the magnetic moments will have to overcome the large energy barrier represented by $\Gamma^{5g}$. Instead, a torque or applied field that induces a chirality-inverting rotation to left-handed $\Gamma^{4g}$, followed by a rotation in (11-1), will relax to the opposite right-handed $\Gamma^{4g}$ state with a smaller energy barrier. In a multi-terminal device this scheme would allow for a separation between read and write currents, with a read current applied along a “hard” direction unable to change the magnetic state, while a write current applied along an “easy” direction will be able to switch the state in an energy-efficient manner.

## Conclusion

In conclusion, we have demonstrated that high-quality single-crystal films of $Mn_3NiN$ are an ideal playground to explore a variety of non-collinear spin configurations with different symmetries and chiralities. The spin-order in these films is sensitive to the coherent thickness, allowing stabilization of the $\Gamma^{4g}$ phase for a wide temperature range at low thickness, extending its operational temperature and maximizing the intrinsic AHE. Increasing the thickness allows a bulk-like first order transition to $\Gamma^{4g}$ and subsequent transition to $\Gamma^{5g}$, and in the thickest films, we see this transition occur by a novel mechanism, with a strikingly sharp, coupled change in the electrical transport properties, magnetic anisotropy and lattice parameters. This result maps a new pathway for energy-efficient switching between octupole nc-AFM orders beyond torques applied in the Kagome plane - which is critical for future spintronic device design. Exploring the dynamics of switching through this pathway will be a component of future work.

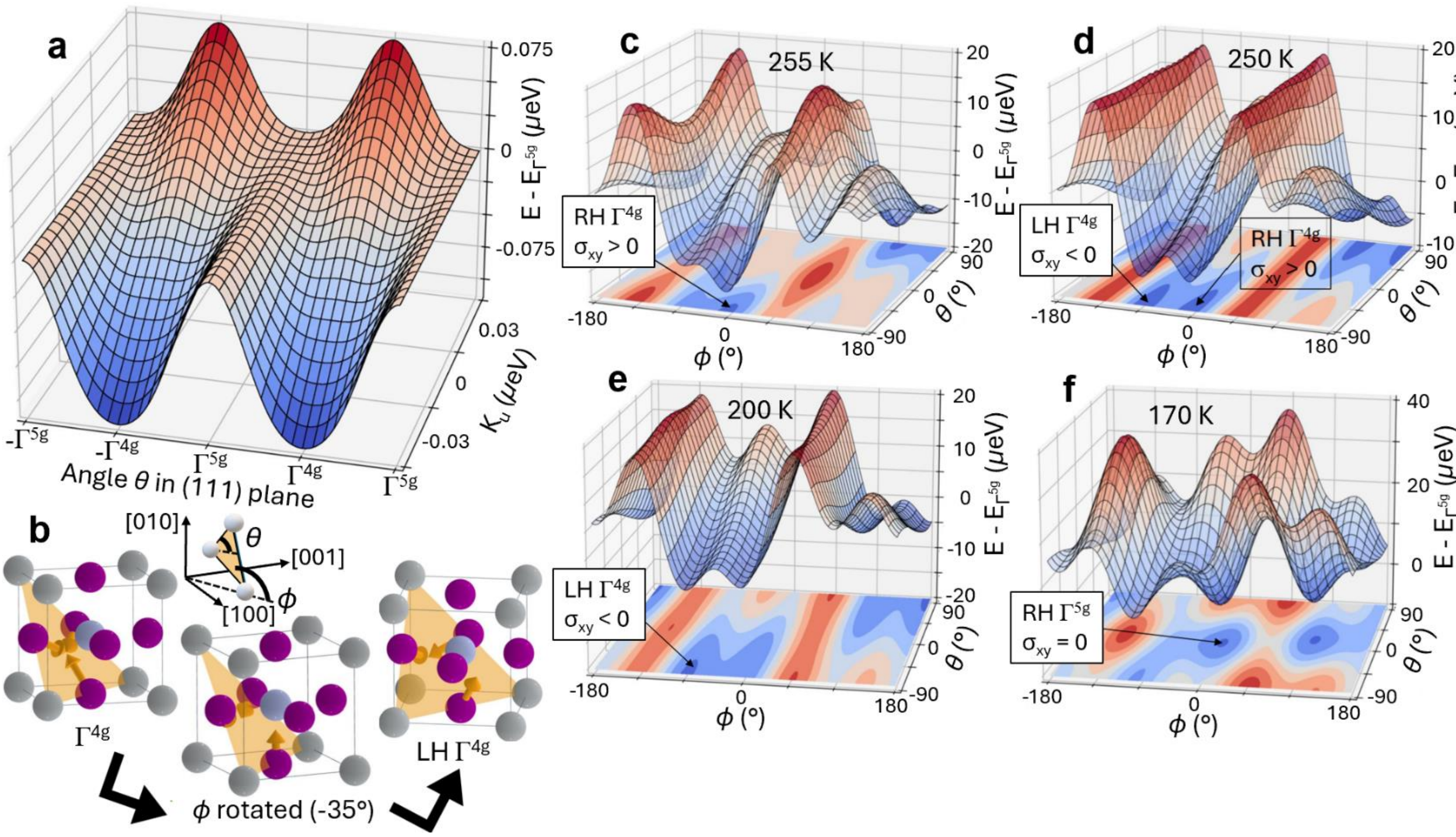

Fig 5. (a) Energy landscape describing the transition from $\Gamma^{4g}$ to $\Gamma^{5g}$ without applied field. The energy is plotted as a function of uniaxial anisotropy, $K_u$, and the spin rotation angle in the (111) plane, $\theta$. (b) Schematic of the alternate rotation sequence around the [110] direction, beginning at the right-handed $\Gamma^{4g}$ ordering and ending at the left-handed $\Gamma^{4g}$ ordering. (c-f) Energy landscape as a function of $\theta$, $\varphi$ under 7 T applied field, with changing anisotropy. (c) Large negative $K_u$ at 255 K favors $\Gamma^{4g}$. (d) Vanishing $K_u$ leads to equal minima for right-handed $\Gamma^{4g}$ and left-handed $\Gamma^{4g}$ at 250 K. (e) Small positive $K_u$ at 200 K favors Type-B. (f) Large positive $K_u$ at 170 K favors $\Gamma^{5g}$.

## Methods

Thin films of $Mn_3NiN$ were grown on (001)-oriented 0.5 mm thick MgO substrates by pulsed laser deposition. The films were deposited at 500 °C under 12 mTorr $N_2$ partial pressure, and then subject to a postgrowth anneal by heating to 700 °C at a rate of 10 °C / min and then cooled to room temperature immediately after reaching the target temperature. The bulk lattice parameter of the $Mn_3NiN$ target used for growth was 3.8805 Å.

Four terminal magnetotransport data were collected using the van der Pauw method, and antisymmetrised to extract the Hall component.

X-ray diffraction (XRD) at room was performed using a Malvern Panalytical Empyrean diffractometer. XRD on cooling was performed using a Rigaku SmartLab Multipurpose X-ray Diffraction System. Software fits of the peaks were performed to extract the lattice parameters.

Magnetic measurements were performed using a Quantum Design Magnetic Property Measurement System SQUID magnetometer. Films were mounted on quartz holders for in-plane measurements and in straws for out-of-plane measurements.

**Details of Macrospin Model**

We model $Mn_3NiN$ as an antiferromagnet composed of three equivalent magnetic sublattices with constant net magnetization. We assume a negligible spatial variation of the magnetic moment within each sublattice in our experimental situation so we can describe our system by three classical vectors $\boldsymbol{M}_1, \boldsymbol{M}_2, \boldsymbol{M}_3$, representing the direction of the three sublattice magnetizations. We adopt the magnetic energy density given by:

$$E_{mag} = -\mu_0 \boldsymbol{H}_{ext} \cdot (\boldsymbol{M}_1 + \boldsymbol{M}_2 + \boldsymbol{M}_3)M$$

$$- J_{12}(M_1 \cdot M_2) - J_{13}(M_1 \cdot M_3) - J_{32}(M_3 \cdot M_2)$$

$$+ \sum_{i=1}^{3} \left(K_{t,z} M_{i,x}^2 M_{i,y}^2 + K_{t,y} M_{i,x}^2 M_{i,z}^2 + K_{t,x} M_{i,y}^2 M_{i,z}^2\right)$$

$$+ K_{u,x} M_{1,x}^2 + K_{u,y} M_{2,y}^2 + K_{u,z} M_{3,z}^2$$

where, $\mu_0$ is the permeability of free space, $H_{ext}$ is the externally applied magnetic field, $M$ is the size of each moment, and $\boldsymbol{M}_i = (M_{i,x}, M_{i,y}, M_{i,z})$ denotes the magnetization unit vector at site $i$ in cartesian components. The Dzyaloshinskii–Moriya interaction (DMI) is expected to be an order of magnitude smaller than the Heisenberg exchange [14] so we omit it in our analysis. We consider only coplanar moments within the so-called exchange plane which is then rotated as a solid body with respect to the crystal lattice:

$M_1 = \left(\cos\left(\frac{2\pi}{3} + \gamma\right), \ \sin\left(\frac{2\pi}{3} + \gamma\right), \ 0\right)$, on site [0,0.5,0.5]

$M_2 = \left(\cos\left(-\frac{2\pi}{3} - \gamma\right),\ \sin\left(-\frac{2\pi}{3} - \gamma\right),\ 0\right)$, on site [0.5,0,0.5]

$M_3 = (1,\ 0,\ 0)$ on site [0.5,0.5,0]

The angle $\gamma$ describes a rotation within the exchange plane that is induced by the lattice strain (piezomagnetic effect) and results in a small net moment that the external field can couple to. In this work, we keep the angle fixed to $\gamma = 0.2^\circ$ in all simulations.

The first term in our energy density is the Zeeman term describing the interaction of the net magnetization with the external field. The Heisenberg exchange terms follow with coupling constants $J_{12}, J_{13}, J_{32}$ that account for symmetric pairwise interactions between magnetic moments. Only nearest neighbor interactions with a negative sign of $J_{ij}$ are considered in this model. This AFM exchange coupling combined with the Kagome lattice in body diagonal planes gives rise to magnetic frustration and strong spin-lattice coupling. However, in this work we focus on magneto-crystalline anisotropies so the effect of strain is modelled only via angle $\gamma$>0, and anisotropy constants, whereas the exchange constants and moment sizes correspond to cubic symmetry ($J_{12} = J_{13} = J_{32}$ = 100 meV/f.u. and $M$ = 2.8 $\mu_B$/Mn, both obtained by fitting to DFT results [25], $|M_1| = |M_2| = |M_3| = 1$). The second-order tetragonal anisotropy terms describe the in-plane angular dependence of each local moment, with anisotropy constants $K_{t,x} = K_{t,y} \neq K_{t,z}$ reflecting the energy cost associated with moment orientation relative to crystal axes. Finally, the site-dependent uniaxial anisotropy is modelled by constants $K_{u,x} = K_{u,y} \neq K_{u,z}$, acting on individual components of the magnetic moments depending on the position of each moment in the unit cell. This comprehensive model enables exploration of complex magnetic configurations in $Mn_3NiN$ including the gradual rotation of all three moments within the (111) plane and rotation of the exchange plane around the face diagonals such as the [1,-1,0] axis. Results for specific combinations of parameters are given in Fig. 5, where energy is plotted as a function of angles $\theta$, $\varphi$ for:

(a) B= 0T, $K_{u,x} = K_{u,y}$ scan, $K_{u,z} = K_{u,x}$*1.1 meV/f.u., $K_{t,x} = K_{t,y}$ = -0.05 meV/f.u., $K_{t,z}$ = -0.055 meV/f.u., switching from $\Gamma^{4g}$ to $\Gamma^{5g}$ .

(c) 255K: $B_z$ = 7T, $K_{u,x} = K_{u,y}$ = -0.005 meV/f.u., $K_{u,z}$ = -0.0055 meV/f.u., $K_{t,x} = K_{t,y}$ = -0.05 meV/f.u., $K_{t,z}$ = -0.055 meV/f.u., $\Gamma^{4g}$ favored.

(d) 250K: $B_z$ = 7T, $K_{u,x} = K_{u,y}$ = -0.00005 meV/f.u., $K_{u,z}$ = -0.000055 meV/f.u., $K_{t,x} = K_{t,y}$ = -0.05 meV/f.u., $K_{t,z}$ = -0.055 meV/f.u., negligible $K_u$ - phases with opposite chirality share equally favored energy minima.

(e) 200K: $B_z$=7T, $K_{u,x} = K_{u,y}$ = 0.001 meV/f.u., $K_{u,z}$ = 0.0011 meV/f.u., $K_{t,x} = K_{t,y}$ = -0.05 meV/f.u., $K_{t,z}$ = -0.055 meV/f.u., crossing into positive $K_u$ but LH $\Gamma^{4g}$ stabilised by $B_z$ = 7T.

(f) 170K: $B_z$=7T, $K_{u,x} = K_{u,y}$ = 0.01 meV/f.u., $K_{u,z}$ = 0.011 meV/f.u., $K_{t,x} = K_{t,y}$ = -0.05 meV/f.u., $K_{t,z}$ = -0.055 meV/f.u., RH $\Gamma^{5g}$ favored.

## TEM Analysis

The 4D-STEM dataset was collected on a Thermo Fisher Spectra in STEM mode by scanning the probe in a 100x100 pixel area with a 0.5 nm step size and collecting a diffraction pattern at each probe position using a direct electron, energy-filtered Gatan K3 camera. The strain maps were generated using the open-source py4DSTEM software by first detecting the diffraction disks using a cross-correlation method and then measuring their shift relative to the median value of these lattice vectors across the film area. [[52]]

## Acknowledgements

F.J. is grateful for support from the Royal Commission of 1851 Research Fellowship. This work was supported by the Henry Royce Institute for Advanced Materials, funded through EPSRC grants EP/R00661X/1, EP/S019367/1, EP/P025021/1 and EP/P025498/1. J.H. and S.F. are supported by JSPS KAKENHI (Grant nos. 24K16999, 24H00039, and 24H02235) and MEXT X-nics (Grant no. JPJ011438). C.C. acknowledges a Royal Society Research Fellowship, a Leverhulme Trust Research Project grant (RPG-2023-271), and a UKRI Frontier Research Guarantee grant (EP/Z000637/1).